\input epsf
\documentstyle[sprocl]{article}
\def\be{\begin{equation}}
\def\ee{\end{equation}}
\def\bea{\begin{eqnarray}}
\def\eea{\end{eqnarray}}

\begin{document}
\title{CELLULAR AUTOMATA AND SELF ORGANIZED CRITICALITY}
\author{MICHAEL CREUTZ}
\address{Physics Department, Brookhaven National Laboratory,
Upton, NY 11973, USA}
\maketitle\abstracts{ Cellular automata provide a fascinating class
of dynamical systems capable of diverse complex behavior.  These
include simplified models for many phenomena seen in nature.  Among
other things, they provide insight into self-organized criticality,
wherein dissipative systems naturally drive themselves to a critical
state with important phenomena occurring over a wide range of length
and time scales.  }

\section{Self-organized Criticality}
Self-organized criticality is a concept aimed at describing a class of
dynamical systems which naturally drive themselves to a state where
interesting physics occurs on all scales\cite{btw}.  The idea provides
a possible ``explanation'' of the omnipresent multi-scale structures
throughout the natural world, ranging from the fractal structure of
mountains, to the power law spectra of earthquake sizes
\cite{mcbak}.  Recent applications include such diverse topics as
punctuated evolution
\cite{pmb} 
and traffic flow \cite{mayanagel}.
The concept has even been invoked to explain the
unpredictable nature of economic systems, i.e. why you can't beat the
stock market \cite{sorin}.  

The prototypical example is a sandpile.  On slowly adding grains of
sand to an empty table, a pile will grow until its slope becomes
critical and avalanches start spilling over the sides.  If the slope
becomes too large, a large catastrophic avalanche is likely, and the
slope will reduce.  If the slope is too small, then the sand will
accumulate to make the pile steeper.  Ultimately one should obtain
avalanches of all sizes, with the prediction of the size for the next
avalanche being impossible without actually running the experiment.

\begin{figure}
\epsfxsize .55\hsize
\centerline {\epsfbox{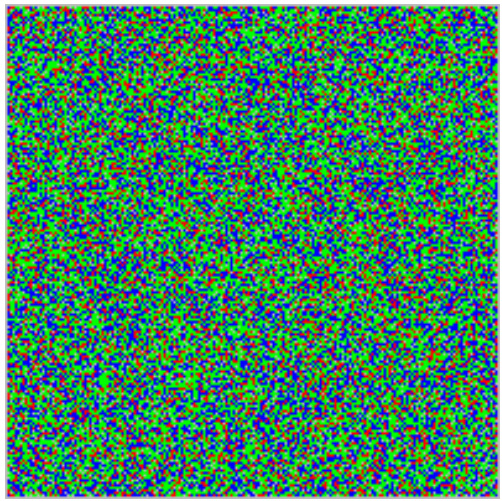}}
\caption{The sandpile model in the final stable state after adding
lots of sand to random places.  The lattice is 198 cells by 198 cells.
The color code is grey, red, blue, and green for heights 0,1,2, and 3,
respectively.  Despite the lack of obvious patterns, subtle
correlations are present; for example no two adjacent sites have
height zero.
\label{fig:sand0}}
\end{figure}

\begin{figure}
\centerline {a.
\epsfxsize .45\hsize \epsfbox{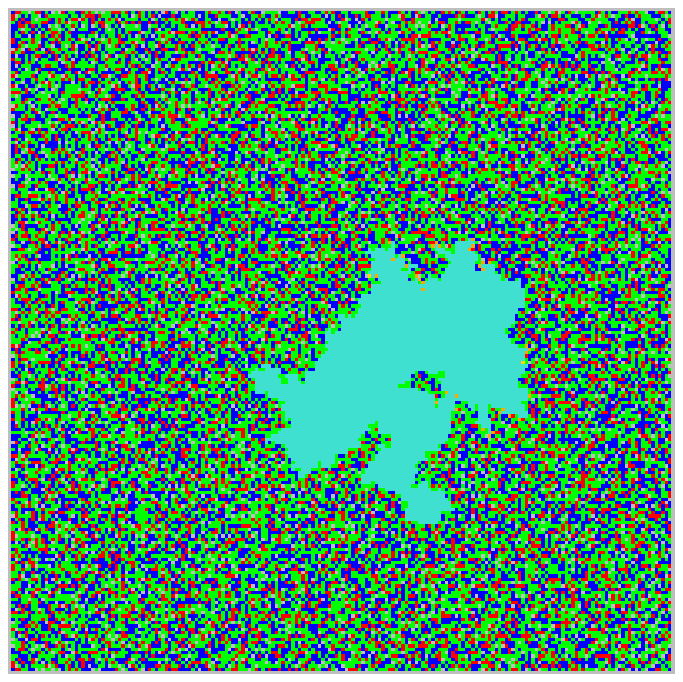}
\hskip .03\hsize 
b.
\epsfxsize .45\hsize \epsfbox{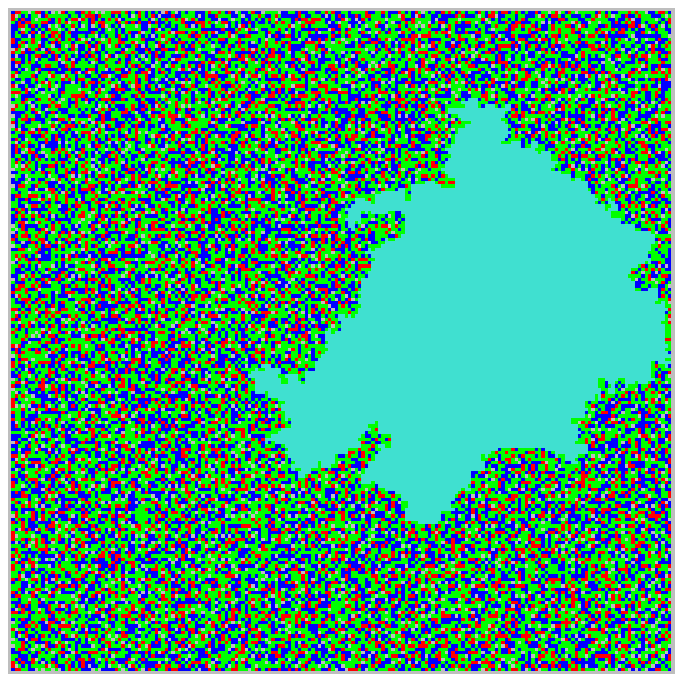}
}
\caption{An avalanche obtained by adding a small amount of sand to the
configuration in Fig.~\protect\ref{fig:sand0}.  Stable sites which
have tumbled during the avalanche are distinguished by being colored
light blue.  The still active sites on the left image are colored
yellowish brown.  The image on the right is the final state after the
avalanche has ended.  Note that the final avalanche region is simply
connected.  This is a general result proven later in the text.}
\label{fig:sand1}
\end{figure}

Self-organized criticality nicely compliments the concept of chaos.
In the latter, dynamical systems with a few degrees of freedom, say
three or more, can display highly complex behavior, including fractal
structures.  With self-organized criticality, we start instead with
systems of many degrees of freedom, and find a few general common
features.  Another attractive feature of this topic is the ease with
which computer models can be implemented and the elegance of the
resulting graphics.  Most of the figures in this chapter were produced
using my publicly available set of programs ``xtoys''\cite{xtoys}.

The original Bak, Tang, Wiesenfeld paper \cite{btw} presented a simple
model wherein each site in a two dimensional lattice has a state
represented by a positive integer $z_i$.  This integer can be thought
of as representing the amount of sand at that location, or in another
sense it represents the slope of the sandpile at that point.  Neither
of these analogies is fully accurate, for the model has aspects of
each.

The dynamics follows by setting a threshold $z_T$ above which any
given $z_i$ is unstable.  Without loss of generality, I take this
threshold to be $z_T=3$.  Time now proceeds in discrete steps.  In one
such step each unstable site ``tumbles'' or ``topples,'' dropping by
four and adding one grain to each of its four nearest neighbors.  This
may produce other unstable sites, and thus an avalanche can ensue for
further time steps until all sites are stable.  Fig.~\ref{fig:sand0}
shows a typical configuration on a 198 by 198 lattice after lots of
random sand addition followed by relaxation.  Fig.~\ref{fig:sand1}
shows an avalanche proceeding on this lattice.

A natural experiment consists of adding a grain of sand to a random
site and measuring the number of topplings and the number of time
steps for the resulting avalanche.  Repeating this many times to gain
statistics, the distribution of avalanche sizes and lengths displays a
power law behavior, with all sizes appearing.  In
Ref.~\cite{christensen} such experiments showed that the distribution
of the number of tumbling events $s$ in an avalanche empirically
scales as
$$
P(s)\sim s^{-1.07}
\eqno(1)
$$
and the number of time steps $\tau$ for avalanches scales as
$$
P(\tau)\sim \tau^{-1.14}
\eqno(2)
$$
This model has been extensively studied analytically.  While as yet
there is no exact calculation of these exponents, a lot is known.  In
particular, the critical ensemble is well characterized.  I will
return to these points later.

The extent to which laboratory experiments reproduce these phenomena
is somewhat controversial.  A recent study of avalanche dynamics
\cite{rice} in
rice piles showed power laws with long-grain rice, but more
ambiguous results followed similar experiments with short-grain rice.

\section{Cellular Automata}

The sandpile model is a simple example of a cellular automaton
system\cite{caref1,caref2}.  Each site or ``cell'' of our lattice follows a
prescribed rule evolving in discrete time steps.  At each step, the
new value for a cell depends only on the current state of itself and
its neighbors.  These systems are fascinating in that deceptively
simple rules can give rise to extremely complex behavior.
Furthermore, slight changes in the rules can dramatically change their
behavior.

Even though the formulation of a cellular automaton may seem almost
trivial, there are a huge number of possible rules.  For example,
suppose I consider two dimensional models where each cell can take
only one of two possible states.  These might be referred to as unset
or set bits, or more figuratively as ``dead'' or ``alive.''  Suppose
furthermore that I restrict myself to rules where the evolution of a
given cell to the next time step depends only on the current values of
the cell and each of its eight nearest neighbors.  In this case there
are $2^9=512$ possible arrangements around a cell, and a general rule
needs to specify the next state for each of these arrangements.  This
gives $2^{512}=1.3\times 10^{154}$ possible rules.  Given that the
universe is only of order $3\times 10^{17}$ seconds old, clearly only
a vanishing fraction of these rules have a chance of being studied in
any of our lifetimes.

A simple subset of rules called ``totalistic'' have the state of the
updated cell only depend on the total number of living neighbors.
With the eight cell neighborhood, there are nine possible values for
this sum, and the new value for the cell requires specification of the
new state for each of these as well as for the current state of the
cell.  This gives $2^{18}=262,144$ rules; still large, but not truly
astronomical.  If I restrict the rule to depending on the total of
only the four nearest neighbors, I then have a modest $2^{10}=1024$
cases to consider.  Other than the sandpile model, most of the
following will be restricted to such totalistic rules.

With a discrete set of states, cellular automata have the appealing
feature of being easily implementable entirely by logical operations,
the natural functions of computer circuitry.  Also, the state of
several cells can be stored and manipulated within a single computer
word.  Using such tricks, these models can often be implemented to run
extremely fast, leading to hope that such models may supply simulation
methods as good as or better than the conventional use of floating
point fields on a discrete grid.  With this motivation, considerable
attention has been paid to cellular automata that may simulate fluid
flow.  Another advantage of this approach is the ability to work with
arbitrary boundary conditions.  These topics go beyond the scope of
this article.  A nice review can be found in Ref.~\cite{bogosian}

\section{Conway's Life}

\begin{figure}
\epsfxsize .5\hsize
\centerline {\epsfbox{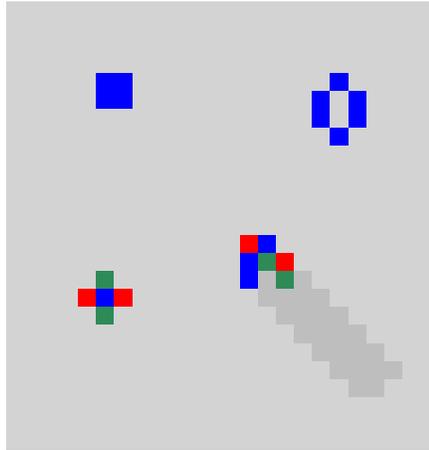}}
\caption{Some living configurations in life.  The top two are stable
patterns.  The lower left shows a ``blinker'' or ``traffic light''
which oscillates with a period of two.  On the lower right is a
glider, which propagates diagonally through the lattice.  Blue denotes
a state that is and just was alive, red is newborn, and green
represents just died.  The track of the glider is darkened slightly
over the remaining grey background.}
\label{fig:life0}
\end{figure}
\begin{figure}
\centerline {
\epsfxsize .45\hsize \epsfbox{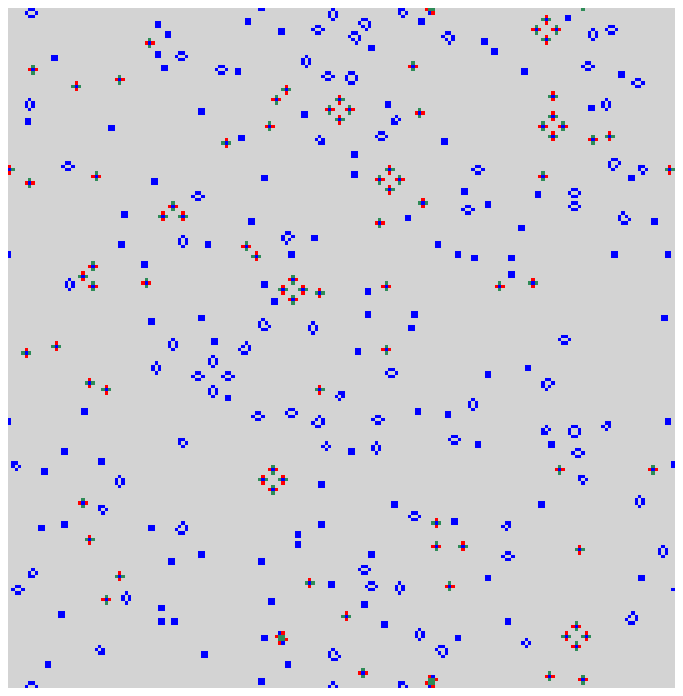}
\hskip .05\hsize
\epsfxsize .45\hsize \epsfbox{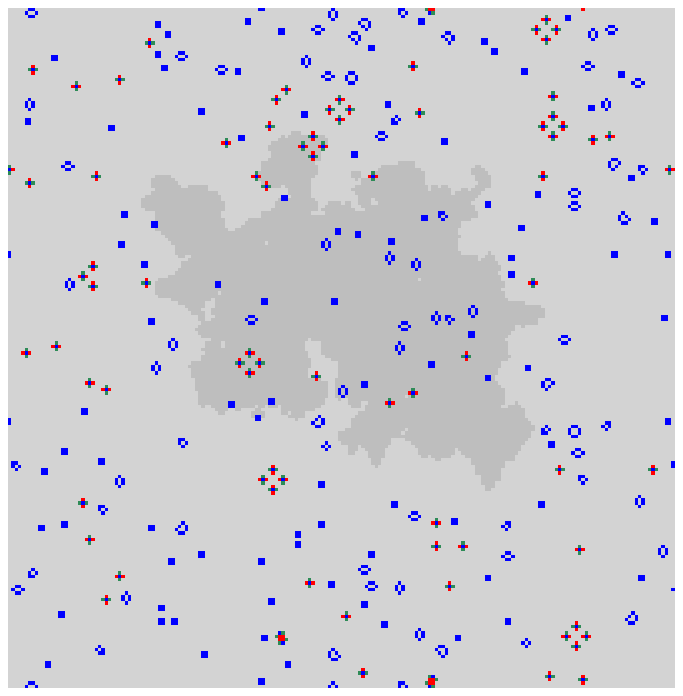}
}
\caption{On the left is a configuration in life resulting from
a random start and evolved until only stable and period two oscillators
remain.  On the right is the state after a small disturbance was introduced
in the center and allowed to die out.  Note the irregular shape of the
disturbed region, which has been tinted a darker grey.  The lattice
here is 198 sites wide by 202 sites high, with periodic boundaries.}
\label{fig:life}
\end{figure}

Perhaps the most famous cellular automaton system is Conway's ``Game
of Life'' \cite{life}.  For this there exists a vast literature; so, I
will only mention a couple of interesting features.  The rule involves
the eight cell neighborhood, and if a cell is initially ``dead'' it
becomes alive if and only if it has exactly three live neighbors, or
``parents.''  A living cell dies of loneliness if it has less than two
live neighbors, and of overcrowding if it has more than three live
neighbors.  Only in the case of exactly two or three live neighbors
does it survive.

While simple to state, this model displays fascinating complexity.
There are simple isolated sets of live cells that quietly survive,
such as a block of four neighboring live cells forming a two by two
square.  Other configurations oscillate, such as three live cells in a
row, which alternate between being vertically and horizontally
oriented.  A particularly amusing local configuration has five live
cells; say starting with coordinates
$\{(0,0),(0,1),(0,2),(1,2),(2,1)\}$.  After four time steps this
configuration returns to its original shape, but displaced by
$(-1,1)$.  On an otherwise empty board, this ``glider'' continues to
propagate as a single entity.  In an on-screen simulation, it appears
much as a small insect crawling about.  Some elementary configurations
are shown in Fig.~\ref{fig:life0}.  A large collection of fascinating
complex life configurations can be found at
\cite{paullife}.

\begin{figure}
\centerline {
\epsfxsize .7\hsize \epsffile{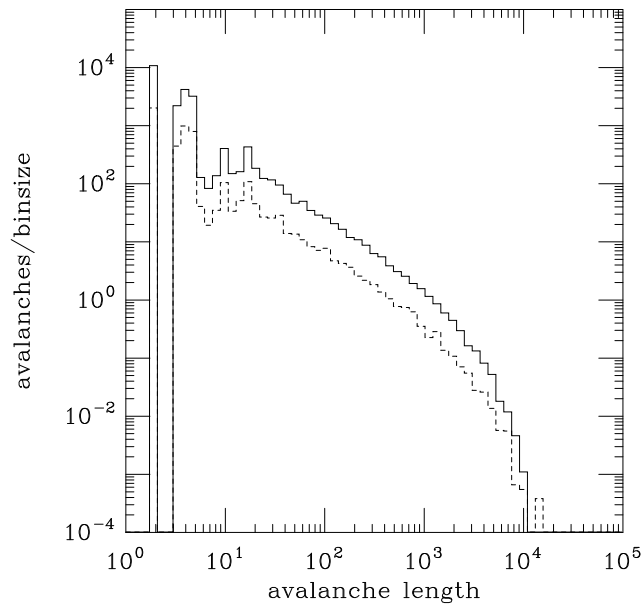}
}
\caption{The distribution of avalanches generated by adding
gliders randomly to a system in the game of life consisting of stable
and period two oscillators.  An avalanche occupies the period until
the system has relaxed again into such a periodic state.  The solid
line represents 25,000 avalanches on a 512 by 512 lattice, and the
dashed line is for 6,000 avalanches on a 1024 by 1024 system.  This
figure is taken from Ref.~\protect\cite{mclat91}.}
\label{fig:lifestats}
\end{figure}

Gliders allow information to be propagated over long distances, and it
has been proven that with a complicated enough initial configuration,
one can construct a computer out of live cells on a life board
\cite{life}.  Special sub-configurations form the analog of
electronic gates, which can control beams of gliders representing
bits.  Indeed, since life is capable of universal computation, one
might imagine a life board programmed to simulate, say the game of
life.

There is some limited evidence that the game of life also displays
self-organized criticality \cite{ourlife,mclat91}.  One can repeatedly
throw down gliders, which collide and create a background of static
and oscillating clumps.  While oscillators of arbitrarily long period
are known to exist, those with period longer than two are extremely
rare and almost never created from unorganized initialization.  Once
the system has settled into a loop, then another glider can be tossed
on, giving a disturbance.  An avalanche is defined to occur during the
period until the system again goes into an oscillating state.
Fig.~\ref{fig:life} shows the effect of such a disturbance.  In
Fig.~\ref{fig:lifestats} I show the distribution of such avalanches as
measured on modest lattices.  There is a hint of a power law
superposed on additional structure from avalanches of only a few time
steps, and a rounding at large times possibly due to finite size
effects.  The criticality of life remains controversial;
Ref.~\cite{bennett} has looked unsuccessfully for a power law
distribution of activity as one moves in from a source on the
boundary.  The relation between these two experiments is unclear.

\section{Fredkin's modulo two rule}

\begin{figure}
\centerline {
\epsfxsize .45\hsize \epsfbox{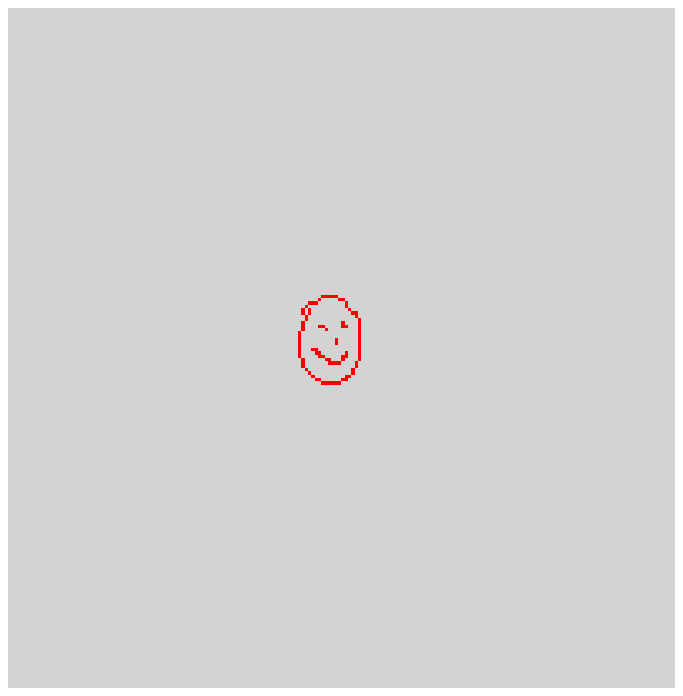}
\hskip .05\hsize 
\epsfxsize .45\hsize \epsfbox{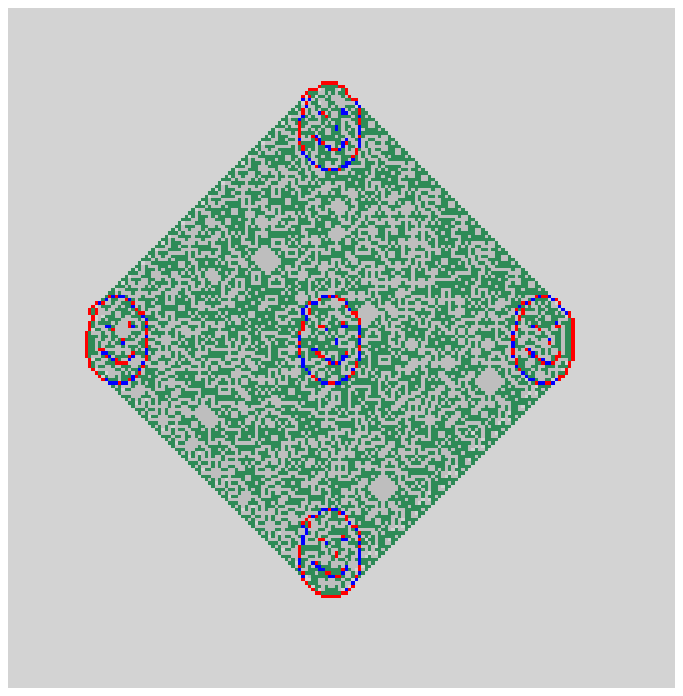}
}
\caption{Starting from the initial configuration on the left,
the modulo two rule is evolved using the four nearest neighbors.  At a
certain stage, five copies of the original image appear.  The green
denotes just expired cells, which form a complex fractal pattern which
has just decayed into the five copies.  }
\label{fig:face}
\end{figure}

A simple but highly amusing rule takes at each time step the
``exclusive or'' (XOR) operation between a site and its neighbors.
This rule has the remarkable property of self replication \cite{xor}.
Starting with any given initial pattern, after $2^n$ time steps copies
of the original state occupy positions separated by $2^n$ spatial
sites from the original in every direction as specified in the chosen
neighborhood.  In Fig.~\ref{fig:face} I show an example of this with
the four cell neighborhood.

\begin{figure}
\centerline {
\epsfxsize .6\hsize \epsfbox{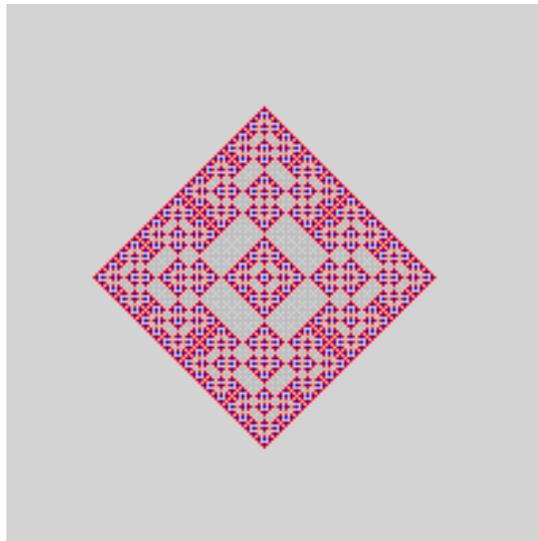}
}
\caption{The state after applying 63 steps of 
the modulo two rule using the four nearest neighbors to an initial
state of a single set bit.  After the next time step this fractal
structure decays into only five remaining live cells.}
\label{fig:xor}
\end{figure}

In this rule, the pattern is generally rather complex just before
returning to the replicated case, i.e. after $2^n-1$ steps.
Fig.~\ref{fig:xor} shows the pattern obtained from a single set
pixel after this rule has been applied for 63 time steps using the
four nearest cells as the neighborhood.  Note the fractal structure.
In one more time step, all but five copies of the original set bit
die.

Unlike most cellular automaton rules, this gives a dynamics which in
some sense is not really ``complex.''  In most cases the simplest way
to predict the evolution of a cellular automaton rule is to actually
run it.  Here, however, I have an easier way to predict what the final
pattern will look like; it is always an XOR operation between several
displaced copies of the configuration that appeared $2^n$ time steps
in the past.  Despite the lack of complexity, this rule shows rather
dramatically that cellular automata are indeed capable of
``reproduction.''

\section{Reversible rules}

Reversibility is rather elusive among cellular automata.  In the game
of life, a single isolated cell immediately dies leaving no trace;
thus it is impossible from the state at a given time to reconstruct
what was there one time step back.  A related difficult problem is to
construct ``garden of Eden'' configurations which are impossible to
arrive at from any previous state.

Fredkin pointed out an interesting class of reversible rules based on
an analogy with molecular dynamics \cite{caref2}.  In the later one
specifies both the position and the velocities of a set of particles
and evolves the system under Newton's equations with some given
inter-particle force law.  Reversal can then be accomplished by merely
changing the signs of all the velocities.  

In a cellular automaton an analog of velocity requires the value of
the cells at two successive time steps.  Based on this, Fredkin
presented a very simple scheme using the previous state to generate a
wide class of reversible rules.  He considered taking an arbitrary
automaton rule at a given time, and then added an exclusive or (XOR)
operation of the result with the state one step back in time.  These
combined operations could then be reversed by merely interchanging two
successive time steps, the analogy of reversing the velocities.

To see this more mathematically, suppose the state at time $i$ is
$s_i$, and the underlying rule begins by taking some arbitrary
function $f(s_i)$.  Then the full rule takes for the next time step
$s_{i+1}=f(s_i)\ {\rm XOR}\ s_{i-1}$.  Here the exclusive or operation
is taken site by site over the entire lattice.  Elementary properties
of the XOR operation then give $s_{i-1}=f(s_i)\ {\rm XOR}\ s_{i+1}$,
which is the identical rule for the time reversed dynamics.

\begin{figure}
\centerline {
a. \epsfxsize .28\hsize \epsfbox{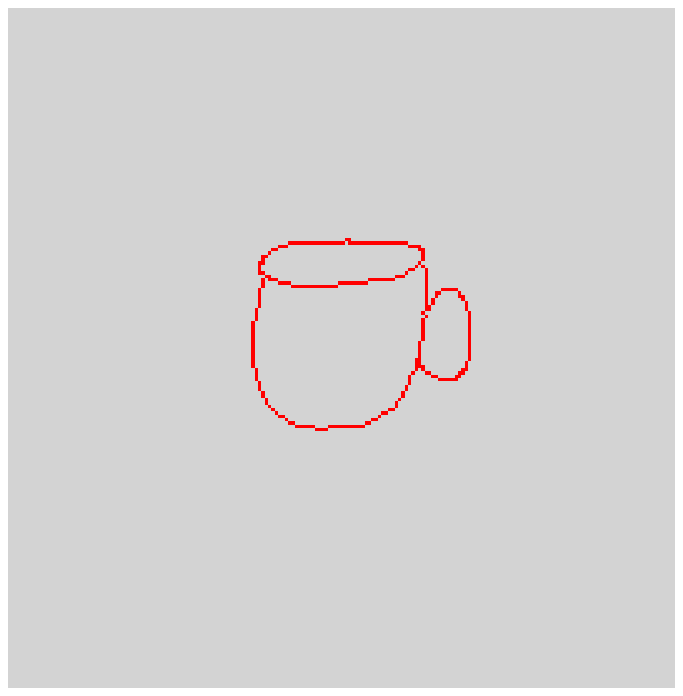}
\hskip .03\hsize 
b. \epsfxsize .28\hsize \epsfbox{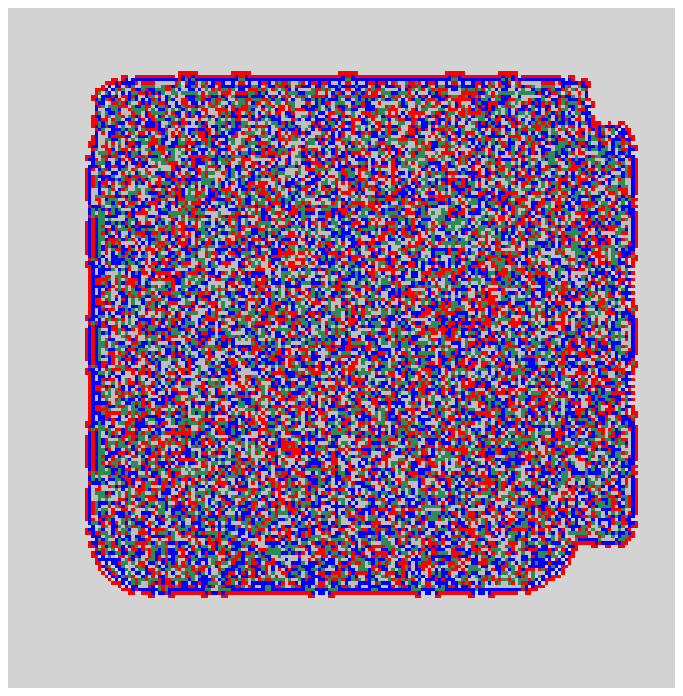}
\hskip .03\hsize
c. \epsfxsize .28\hsize \epsfbox{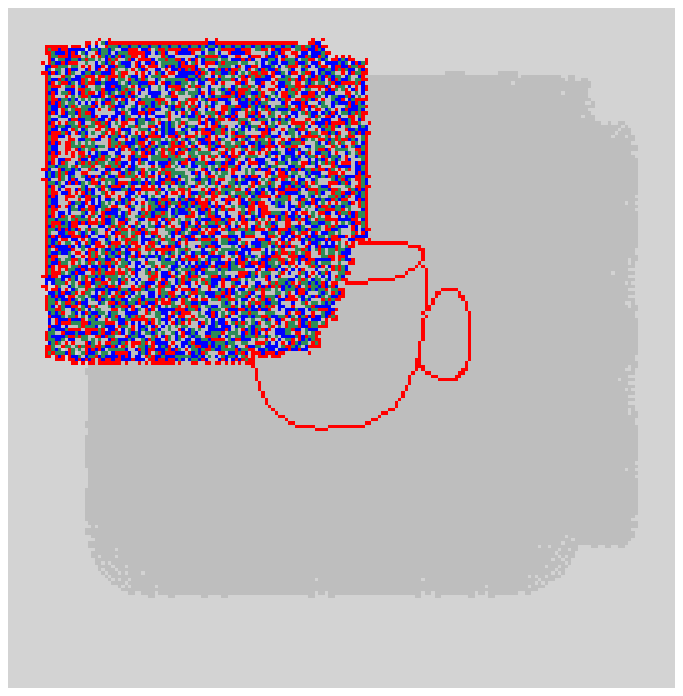}
}
\caption{The encryption of a coffee cup.  The original rule uses the
eight cell neighborhood with births on 1,3,4, and 7 neighbors and
survivors on 0,2,3,4, and 8 neighbors.  The rule is modified by XOR'ing
the result with the cell one time step back.  Swapping two adjacent time
steps will bring the encrypted coffee cup back exactly.  The third image
is what happens if you reverse the process except make an error on one
bit in the upper left hand quadrant.  Note the effect of a ``speed of
light'' in the problem.}
\label{fig:coffee}
\end{figure}

These rules provide a wonderful way to play with the concepts of
entropy and reversibility.  Indeed, an idealized universe of cellular
automata enables experiements which would be impossible to carry out
in the real world.  In Fig.~\ref{fig:coffee} I show the evolution of a
simple image under such a rule.  The experiment is a crude simulation
of a coffee cup shattering after being dropped on the floor.  After a
few steps it appears quite randomized.  Reversal of the momenta of all
relevant atoms in the coffee cup would allow its reconstruction.  In
the model this is easily accomplished by swapping two time steps.
After reversal, continuing with the same rule reconstructs the
original image.  At all stages the ``information'' contained in the
system must be constant, even though the image may appear of
drastically different complexity.

\begin{figure}
\hbox{
a.
\epsfxsize .45\hsize
\epsffile{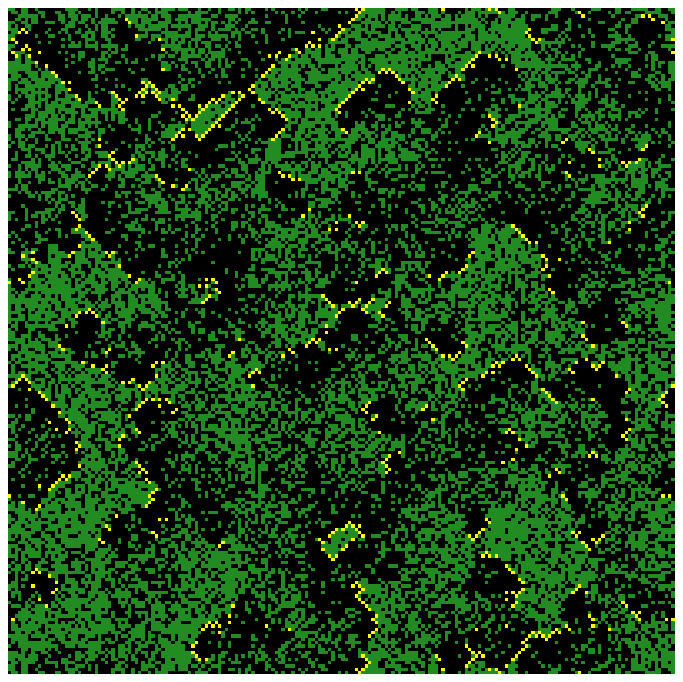}
\hskip .02\hsize
b.
\epsfxsize .45\hsize
\epsffile{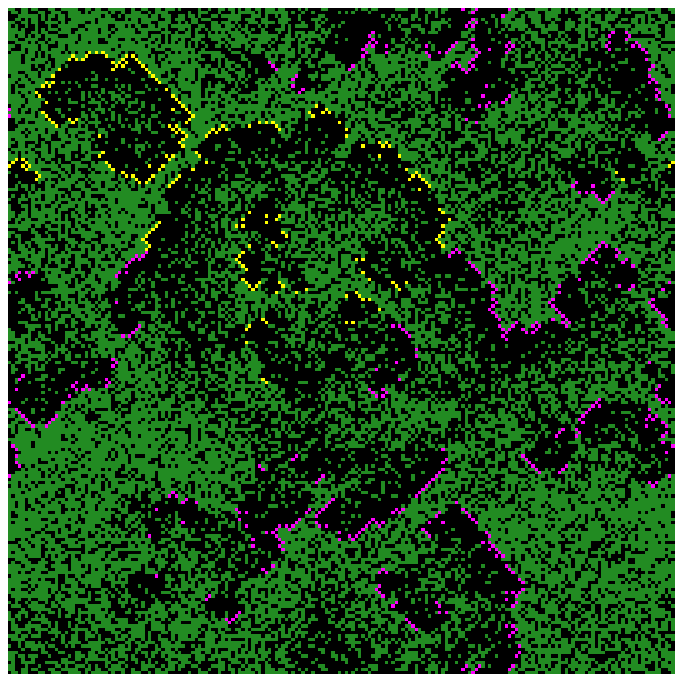}
} 
\caption{On the left is a snapshot of the forest fire model on a 198 by 198
lattice.  Trees are continuously burning at a slow rate, while fires
burn them down and spread to nearest neighbor trees.  Here the four
cell neighborhood is used.  On the right is a variation where two
species of bunnies are competing to eat a common grass.  The yellow
and purple colors here distinguish the parity of the site plus time
step.  Eventually one of the two species dominates and the other dies
out.}
\label{fig:fires}
\end{figure}

The reconstruction process is highly sensitive to the reversal being
precise.  The analog here is to the sensitivity to initial conditions
in dynamical systems.  In Fig.~\ref{fig:coffee}c I try to
reproduce the coffee cup from its shards as in the above experiment,
except that now at the time of reversal I modify the state of exactly
one pixel.  The reversal process recovers the original image only in
regions outside the ``light cone'' for the modified pixel.  As the
disturbance can only propagate to neighbors in one time step, pixels
outside $n$ steps can not know of the change before an equal number of
time steps.  This use of an XOR operation to generate reversible
complex mappings is an integral part of the Data Encryption Standard;
see, for example, Ref.~\cite{recipes}.

\section{Forest fires and bunny wars}

An amusing model of forest fires has three possible states per cell,
empty, a tree, or a fire.  For the updating step, any empty site can
have a tree born with a small probability.  At the same time, any
existing fire spreads to neighboring trees leaving its own cell empty.
The rule here differs from those discussed previously in having a
stochastic nature.  As the system is made larger, the growth rate for
the trees should decrease to just enough to keep the fires going.

If too many trees grow, one obtains a large fire reducing their
density, while if there are too few trees, fires die out.  On a finite
system, one should light a fire somewhere to get the system started.
On the other hand, as the system becomes larger, the growth rate for
the trees can be reduced without the fire expiring.  In a steady state
the system has fire fronts continually passing through the system, as
illustrated in Fig.~{\ref{fig:fires}a.}.  Perhaps there is a moral
here that one should be careful about extinguishing all fires in the
real world, for this may enhance the possibility for a catastrophic
uncontrollable fire.  It is not entirely clear whether this model is
actually critical.  What seems to happen on large systems is that
stable spiral structures form and set up a steady rotation.  For a
review of this and several related models, see Ref.~\cite{cds}.

A variation on this model has several ``species'' of fires.  Perhaps a
better metaphor is to think of different species of bunny, competing
for the same slowly growing food resource.  With the four cell
neighborhood, a natural division into species is given by the parity
of the site plus the time step.  Fig.~{\ref{fig:fires}b.}  shows a
state in the evolution of such a model when both species are present.
This situation, however, is highly unstable, with any fluctuation
favoring one species tending to grow until the competitor is
eliminated.  This model provides a discrete realization of the
``principle of competitive exclusion'' in biological systems
\cite{murray}.  Stability of a species requires that it occupy its own
niche and not compete for exactly the same resources as another.

\section{The sandpile revisited}

Very little is known analytically about general cellular automata.
However, in a series of papers, Deepak Dhar and co-workers have shown
that the sandpile model has some rather remarkable mathematical properties
\cite{dar1,dar2,dar3,dar4}. In particular, the critical ensemble of 
the system has been well characterized in terms of an Abelian group.
In the following I will generally follow the discussion given in
Refs.~\cite{mysand,mcbak}.  

Dhar introduced the useful toppling matrix $\Delta_{i,j}$ with integer
elements representing the change in the height, $z$ at site $i$
resulting from a toppling at site $j$ \cite{dar1}. More precisely,
under a toppling at site $j,$ the height at any site $i$ becomes $z_i -
\Delta_{i,j}$.  For the simple two dimensional sand model the toppling
matrix is thus
$$ \matrix
{
\Delta_{i,j}&= 4    &   i=j     \cr               
\Delta_{i,j} &=-1   & i, j {\rm \ \ nearest\ neighbors}     \cr 
\Delta_{i,j} &= 0   &        {\rm \ \ otherwise.}   \cr
}
\eqno (3)
$$

For this discussion there is little special to the specific lattice
geometry; indeed, the following results easily generalize to other
lattices and dimensions.  The analysis requires only that under a
toppling of a single site $i,$ that site has its slope decreased
$(\Delta_{i,i} > 0)$, the slope at any other site is either increased
or unchanged $(\Delta_{i,j} \leq 0, j \ne i)$, the total amount of
sand in the system does not increase $(\sum_j \Delta_{i,j} \ge 0)$, and,
finally, that each site be connected through toppling events to some
location where sand can be lost, such as at a boundary.

For the specific case in Eq. 3, the sum of slopes over all sites is
conserved whenever a site away from the lattice edge undergoes a
toppling. Only at the lattice boundaries can sand be lost. Thus the
details of this model depend crucially on the boundaries, which we
take to be open.  A toppling at an edge loses one grain of sand and at
a corner loses two.

The actual value of the maximum stable height $z_T$ is unimportant to
the dynamics.  This can be changed by simply adding constants to all
the $z_i$. Thus without loss of generality I consider $z_T = 3$.  With
this convention, if all $z_i$ are initially non-negative they will
remain so, and I thus restrict myself to states $C$ belonging to that
set. The states where all $z_i$ are positive and less than 4 are
called stable; a state that has any $z_i$ larger than or equal to 4 is
called unstable. One conceptually useful configuration is the
minimally stable state $C^*$ which has all the heights at the critical
value $z_T$.  By construction, any addition of sand to $C^*$ will give
an unstable state leading to a large avalanche.

I now formally define various operators acting on the states
$C$. First, the ``sand addition" operator $\alpha_i$ acting on any $C$
yields the state $\alpha_iC$ where $z_i = z_i+1$ and all other $z$ are
unchanged.  Next, the toppling operator $t_i$ transforms $C$ into the
state with heights $z^\prime_j$ where $z^\prime_j = z_j -
\Delta_{i,j}$. The operator $U$ which updates the lattice one time
step is now simply the product of $t_i$ over all sites where the slope
is unstable,
$$
UC=\prod_i t_i^{p_i}C
\eqno(4)
$$
where $p_i = 1$ if $z_i\geq 4$; $0$ otherwise.  Using $U$ repeatedly
gives the relaxation operator $R$. Applied to any state $C$ this
corresponds to repeating $U$ until no more $z_i$ change.  Neither $U$
nor $R$ have any effect on stable states.  Finally, I define the
avalanche operators $a_i$ describing the action of adding a grain of
sand followed by relaxation
$$
a_iC = R\alpha_iC. \eqno (5)
$$

At this point it is not entirely clear that the operator $R$ exists;
in particular, it might be that the updating procedure enters a
non-trivial cycle consisting of a never ending avalanche. I now prove
that this is impossible.  First note that a toppling in the interior
of the lattice does not change the total amount of sand. A toppling on
the boundary, however, decreases this sum due to sand falling off the
edge. Thus, during an avalanche the total sand in the system is a
non-increasing quantity. No closed cycle can have toppling at the
boundary since this will decrease the sum.  Next, the sand on the
boundary will monotonically increase if there is any toppling one site
further in.  This also can not happen in a cycle; thus, there can be
no topplings one site away from the edges. By induction there can be
no toppling arbitrary distances in from the boundary; thus, there can
be no cycle, and the relaxation operator exists. Note that for a
general geometry this argument requires that every site be eventually
connected to an edge where sand can be lost. 

With an edge less system, such as under periodic boundaries, no sand
would be lost and thus cycles are expected and easily observed. These
models might be called ``Escher models" after the artist constructing
drawings of water flowing perpetually downhill and yet circulating in
the system.  While little is known about the dynamics of this
variation on the sandpile model, some studies have been done under the
nomenclature of ``chip-firing games'' \cite{chipfiring}.  A recent
paper \cite{unisand} has argued that this lossless sandpile model on
an appropriate lattice is capable of universal computation.

I now introduce the concept of recursive states. This set, denoted
${\cal R}$, includes those stable states which can be reached from any
stable state by some addition of sand followed by relaxation.  This
set is not empty because it contains at least the minimally stable
state $C^*$.  Indeed, that state can be obtained from any other by
carefully adding just enough sand to each site to make $z_i$ equal to
three.  Thus, one might alternatively define ${\cal R}$ as the set of
states which can be obtained from $C^*$ by acting with some product of
the operators $a_i$.

It is easily shown that there exist non-recursive, transient states;
for instance, no recursive state can have two adjacent heights both
being zero.  If you try to tumble one site to zero height, then it
drops a grain of sand on its neighbors.  If you then tumble a neighbor
to zero, it dumps a grain back on the original site.  One can also
show that the self-organized critical ensemble, reached under random
addition of sand to the system, has equal probability for each state
in the recursive set.  This is a consequence of the Abelian nature
of this system, as discussed below.

The crucial results of Refs.~\cite{dar1,dar2,dar3,dar4} are that the
operators $a_i$ acting on stable states commute, and they generate an
Abelian group when restricted to recursive states. I begin by showing
that the operators commute, that is $a_ia_jC = a_ja_iC$ for all
$C$. First I express the $a$'s in terms of toppling and adding
operators
$$
a_ia_jC=\left( \prod_{k=1}^{n_1} t_{l_k} \right )
\alpha_i \left(  \prod_{k=n_1+1}^{n} t_{l_k} \right )
\alpha_j C \eqno (6)
$$
where the specific number of topplings $n_1$ and $n$ depend on $i,$
$j,$ and $C$.  Acting on general states, the operators $t$ and
$\alpha$ all commute because they merely linearly add or subtract
heights. Therefore I can shift $\alpha_i$ to the right in this
expression:
$$
a_ia_jC=\left(  \prod_{k=1}^{n} t_{l_k} \right )
\alpha_i \alpha_j C \eqno (7)
$$
Now I rearrange the product of topplings. In the non-trivial case
that the $\alpha$-operators render either $i$ or $j$ (or both)
unstable, the product must contain toppling operators corresponding to
those unstable sites.  I shift those operators to the right. Those
operators constitute by definition the update operator, U, so I can
write
$$
a_ia_jC=\left(  \prod t_{l_k} \right )                    
U\alpha_i \alpha_j C \eqno (8)
$$

The factors within the bracket are the remaining $t$'s. Now, the
update operator may leave some sites still unstable, and then the
product must include further toppling operators; working on those
sites, I can pull out another factor of the update operator.  This
procedure can be repeated until I have used all the toppling factors
and the state is stable. Thus, I can identify the operator within the
brackets in Eq. (8) as the relaxation operator $R$. But
$\alpha_i\alpha_j C$ is the same state as $\alpha_j\alpha_i C$, so
$a_ia_jC = a_ja_iC$.

A trivial consequence of this argument is that the total number of
tumbling events occurring in the operations $a_ia_jC$ and $a_ja_iC$
are the same. Of course, if a particular site $k$ tumbles it can be
caused by either addition; the orders of the tumbling events may or may not
be altered.

An intuitive argument that sand addition may be commutative uses an
analogy with combining many digit numbers under long addition.  The
tumbling operation is much like carrying, except rather than to the
next digit the overflow spreads to several neighbors.  As addition is
known to be Abelian, despite the confusing elementary-school rules, I
might expect the sandpile addition rule also to be.

I now prove that the avalanche operators have unique inverses when
restricted to recursive states; that is, there exists a unique
operator $a_i^{-1}$ such that $a_i(a_i^{-1}C) = C$ for all $C$ in
$\cal R$. This implies that the operators $a_i$ acting on the
recursive set generate an Abelian group. For any recursive state $C$
I first find another recursive state such that $a_i$ acting on it
gives $C$, and I then show that this construction is unique.

I begin by adding a grain of sand at site $i$ to the state $C$ and
then relax the system. This generates a new recursive state $a_i
C$. Now since the state $C$ is by assumption recursive, there is some
way to add sand to regenerate $C$ from any given state.  In
particular, there is some product $P$ of addition operators $a_j$ such
that
$$
C=P a_i C \eqno (9)
$$
But the $a$'s commute, so I have 
$$
C=a_i PC \eqno (10)
$$
and thus $PC$ is a recursive state on which $a_i$ gives $C$. 

I must now show that this state is unique.  Consider repeating the
above process to find a series of states $C_n$ satisfying
$$
(a_i)^nC_n = C. \eqno (11)
$$
Because on a finite system the total number of stable states is
finite, the sequence of states $C_n$ must eventually enter a loop.  I
can run backwards around this loop by adding back the sand repeatedly
to the given site.  As the original state $C$ appears in resupplying
the sand, $C$ itself must itself belong to the loop. Calling the
length of the loop $m$, I have $(a_i)^mC = C$.  I now uniquely define
$a_i^{-1}C = a_i^{m-1}C$.

I now have sufficient machinery to count the number of recursive
states.  As all such can be obtained by adding sand to $C^*$ , I can
write any state $C\in {\cal R}$ in the form
 $$
 C=\left(\prod_i a_i^{n_i} \right)C^*.    \eqno (12)
 $$
Here the integers $n_i$ represent the amount of sand to be added at
the respective sites. However, in general there are several different
ways to reach any given state. In particular, adding four grains of
sand to any one site must force a toppling and is equivalent to adding
a single grain to each of its neighbors. This can be expressed as the
operator statement
 $$
 a_i^4=\prod_{j\in nn} a_j \eqno (13)
 $$       
where the product is over the nearest neighbors to site $i$.  I can
rewrite this equation by multiplying by the product of inverse
avalanche operators on the nearest neighbors on both sides, thus
obtaining for any site $i$
$$
\prod_j a_j^{\Delta_{ij}} = E \eqno (14)
$$
where $E$ is the identity operator. This allows me to shift the powers
appearing in Eq. (12).  Define $N$ to be the number of sites in the
system.  If I label states by the vector ${\bf n} = (n_1, n_2,
n_3,\ldots n_N)$ I see that two states are equivalent if the
difference of these vectors is of the form $\sum_j\beta_j\Delta_{ij}$
where the coefficients $\beta_j$ are integers. These are the only
constraints; if two states can not be related by toppling they are
independent. Thus any vector ${\bf n}$ can be translated repeatedly
until it lies in an $N$-dimensional hyper-parallelepiped whose base
edges are the vectors $\Delta_{ji}$, $j=1, \ldots N.$ The vertices
of this object have integer coordinates and its volume is the number
of integer coordinate points inside it. This volume is just the
absolute value of the determinant of $\Delta$ .  Thus the number of
recursive states equals the absolute value of the determinant of the
toppling matrix $\Delta$.

For large lattices this determinant can be found easily by Fourier
transform. In particular, whereas there are $4N$ stable states, there
are only
$$
   \exp\left( N\int_{(-\pi,-\pi)}^ {(\pi,\pi)} {d^2q\over (2\pi)^2}  
   \ln(4-2q_x-2q_y)\right)     \simeq (3.2102..)^N     
   \eqno (15)           
$$
recursive states.  Thus starting from an arbitrary state and adding
sand, the system ``self-organizes" into an exponentially small subset
of states forming the attractor of the dynamics.
 
\section {An isomorphism}

Following Ref.~\cite{mysand}, I now look into the consequences of stacking
sand piles on top of one another. Given stable configurations $C$ and
$C^\prime$ with configurations $z_i$ and $z_i^\prime$, I define the
state $C\oplus C^\prime$ to be that obtained by relaxing the
configuration with heights $z_i + z_i^\prime$.  Clearly, if either $C$
or $C^\prime$ are recursive states, so is $C\oplus C^\prime$.
 
Under the operation $\oplus$ the recursive states form an Abelian
group isomorphic to the algebra generated by the $a_i$. First, the
addition of a state $C$ with heights $z_i$ is equivalent to operating
with a product of $a_i$ raised to $z_i$, that is
$$
B \oplus  C = \left(\prod a_i^{z_i}\right)B. \eqno (16)
$$
The operation $\oplus$ is associative and Abelian because the
operators $a_i$ are.

Since any element of a group raised to the order of the group gives
the identity, it follows that $a_i^{|\Delta|} = E$.  This implies the
simple formula $a_i^{-1}= a_i^{|\Delta|-1}$.  The analog of this for
the states is the existence of an inverse state, -$C$
$$
-C = (|\Delta|-1) \otimes C. \eqno (17)
$$
Here, $n \otimes C$ means adding $n$ copies of $C$ and relaxing.  The
state -$C$ has the property that for any state $B\oplus C\oplus(-C) =
B$.

The state $I = C\oplus(-C)$ represents the identity and has the
property $I \oplus B = B$ for every recursive state $B$.  The state
which is isomorphic to the operator $a_i$ is simply $a_iI$. The
identity state provides a simple way to check if a state, obtained for
instance by a computer simulation, has reached the attractor, i.e. if
a given state is a recursive state: A stable state is in ${\cal R}$ if
and only if $C\oplus I = C$. The proof is simple. By construction, a
recursive state has this property. On the other hand, since $I$ is
recursive, so is $C\oplus I$.

\begin{figure}
\epsfxsize \hsize
\centerline {\epsfbox{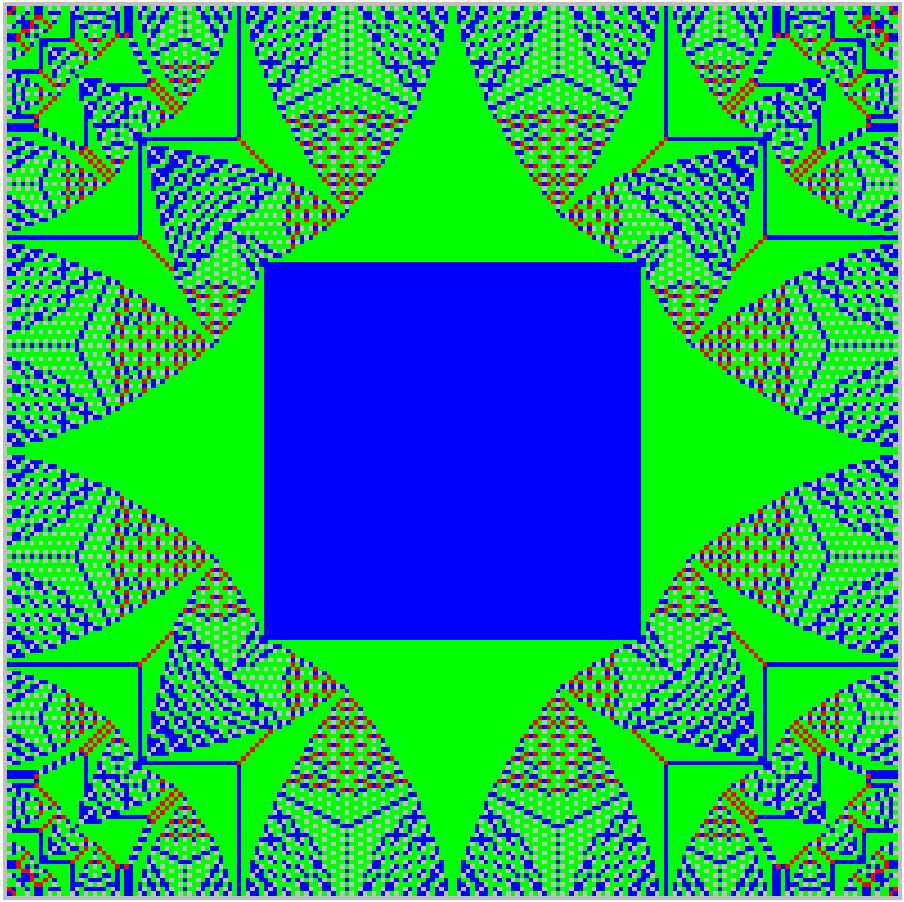}}
\caption{The identity state for the sandpile model on a 198 by 198 lattice.
The color code is grey, red, blue, and green for heights 0,1,2, and 3,
respectively.  
}
\label{fig:identity}
\end{figure}

\begin{figure}
\centerline
{a.
\epsfxsize .45\hsize \epsfbox{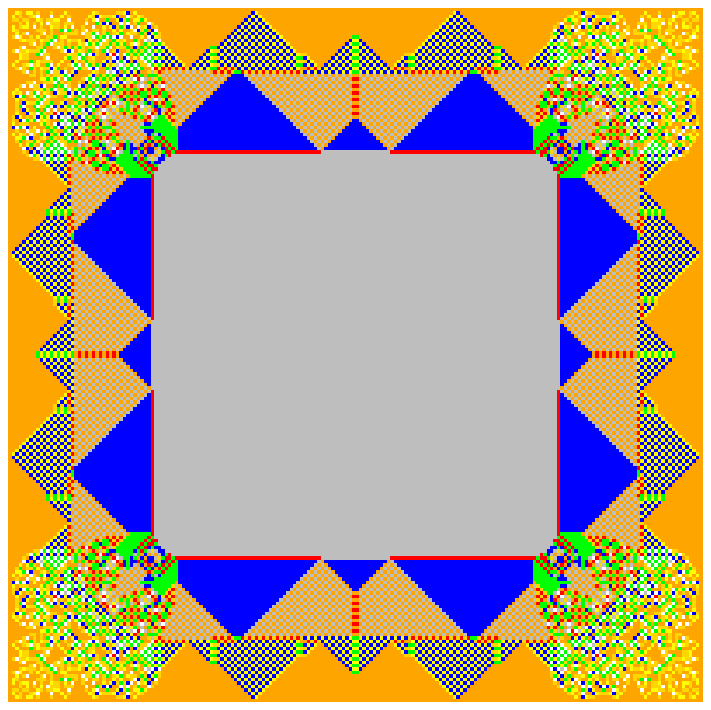}
\hskip .02\hsize 
b.
\epsfxsize .45\hsize \epsfbox{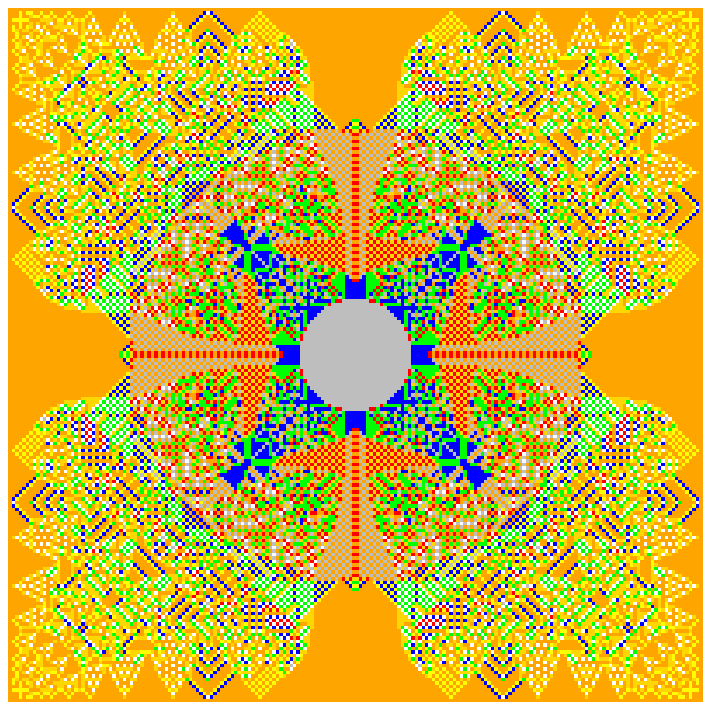}
}
\vskip .2 in 
\centerline {c.
\epsfxsize .45\hsize \epsfbox{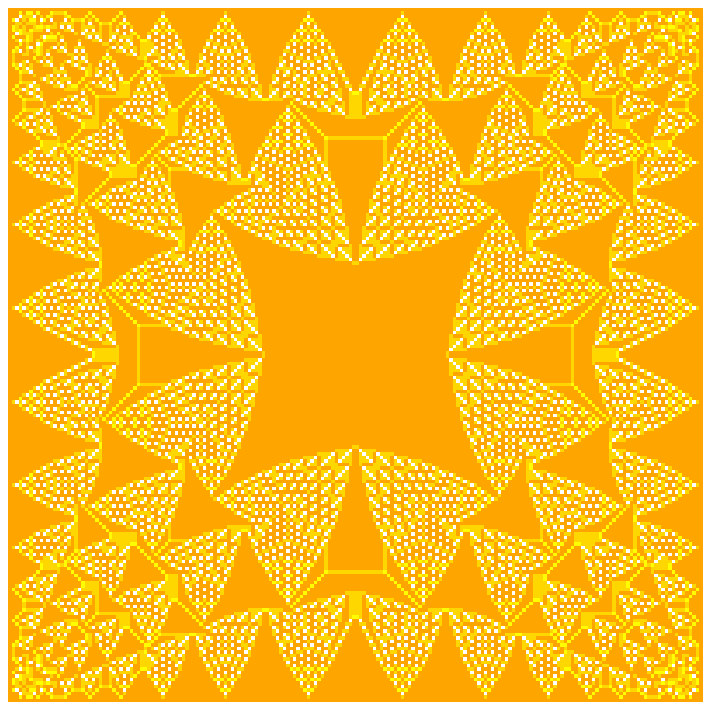}
\hskip .02\hsize 
d.
\epsfxsize .45\hsize \epsfbox{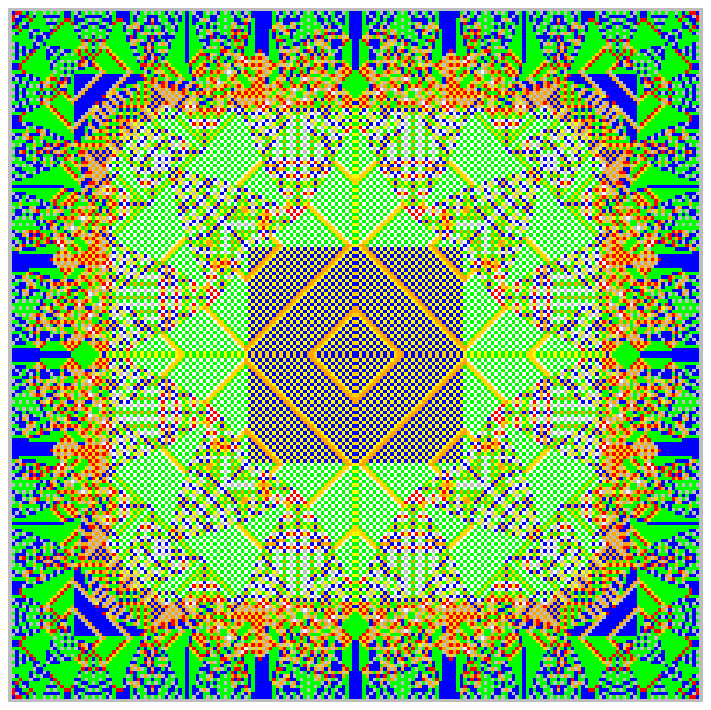}
}
\caption{
A procedure for constructing the identity.  In parts a and b an empty
table is having sand poured on from the edges.  In part c the entire
table is supercritical.  In part d the boundaries have been opened,
and the sand is running back off.  The system finally relaxes the
state shown in Fig.~\protect\ref{fig:identity}.  In this sequence,
heights are color coded as in Fig.~\protect\ref{fig:sand0}, with
active sites in various shades of orange.  The lattice is 198 sites
wide by 198 sites high.}
\label{fig:ident}
\end{figure}

 The identity state can be constructed by taking any recursive state,
say $C^*$ and repeatedly adding it to itself to use $|\Delta| \otimes C =
I$.  However, on any but the smallest lattices, $|\Delta|$ is a very
large integer.  A more economical scheme is given in
Ref.~\cite{mysand}.  Fig.~\ref{fig:identity} shows the identity state
on a 198 by 198 lattice.  Note the fractal structure, with features on
many length scales.

Here I present another way to construct this state.
Fig.~\ref{fig:ident}a-b shows a sequence of configurations obtained by
pouring sand in from the boundaries onto an initially empty table.
This is accomplished by adding a temporary set of sites just outside
the boundary and keeping their heights always supercritical.  A
variety of fractal structures emerge as the interior slowly fills
up. Fig.~\ref{fig:ident}c shows the final stationary state where the sand
falling in from the boundaries matches that falling off from the
updating.  Then I revert the boundary conditions to open and allow the
sand to fall back off.  Fig.~\ref{fig:ident}d shows an intermediate state of
this procedure.  When the sand finally stops falling off, I obtain the
identity state as shown in Fig.~\ref{fig:identity}.

\begin{figure}
\centerline {\epsfxsize \hsize \epsfbox{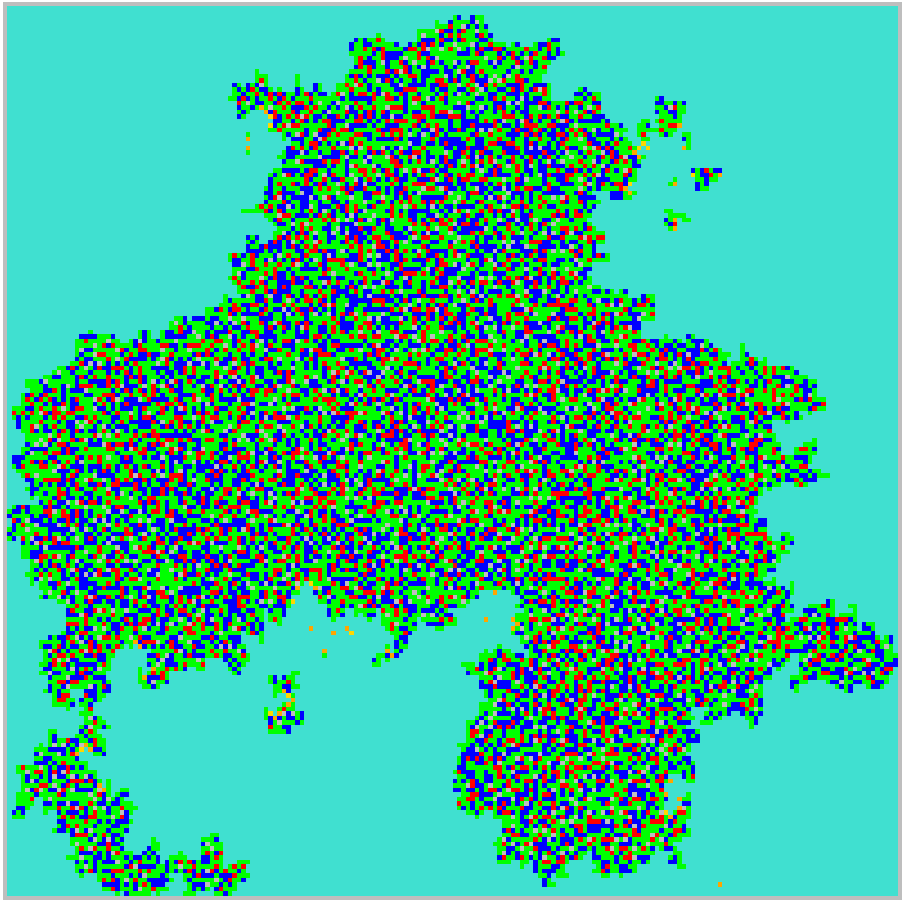}
}
\caption{
The burning algorithm being applied to the state in
Fig.~\ref{fig:sand0}.  Burnt sites are cyan, burning sites are orange,
and the remaining sites are colored as previously.  This avalanche
eventually tumbles every site exactly once.}
\label{fig:burning}
\end{figure}

Majumdar and Dhar \cite{dar4} have constructed a simple ``burning''
algorithm to determine if a state belongs to the recursive set.  For a
given configuration, first add one particle to each of the edge sites
and two particles to the corners.  This again corresponds to imagining
a large source of sand just outside the boundaries, which then tumbles
one step onto the system.  Then return to open boundaries and update
according to the usual rules.  If and only if the original state is
recursive, this will generate an avalanche under which each site of
the system tumbles exactly once.  Also, the final state after the
avalanche will be identical to the original.  However, if the state is
not recursive, some untumbled sites will remain.
Fig.~\ref{fig:burning} shows such a process underway on the
configuration of Fig.~\ref{fig:sand0}.  Here sites which have already
burned are shown in cyan, while the remaining sites in the center have
not yet tumbled.  The small number of sites shown in orange are the
still active sites, which eventually burn the entire remaining
lattice.

The burning algorithm provides a simple way to prove that the
avalanche regions are simply connected once one is in the critical
state.  In a burning process, any sub-lattice of the original will
have all of its sites tumbled onto from outside.  This is the
condition for starting a burning on the sub-lattice.  Thus, if a
configuration is in the critical ensemble for the whole lattice, then
any extracted piece of this configuration on a subset of the original
lattice is also in the critical ensemble of the extracted part.  Now
suppose that one constructs an avalanche with any initial addition to
a state from the critical ensemble.  In any subregion enclosed by this
avalanche, sand will fall from the tumbling sites on its outside.
Since the sub-lattice is itself in its own critical ensemble, this
must induce an avalanche which, by the burning algorithm, will tumble
all enclosed sites.  Thus any avalanche on a state from the critical
ensemble cannot leave untumbled any sites in a region isolated from
the boundary, i.e. an untumbled island.  This result that avalanches
must be simply connected does not follow for states outside the
recursive set, as can be easily demonstrated by considering a sandpile
with a hole of empty sites in the middle.

\section{Concluding remarks}
Simple models as implemented by cellular automata provide a rich area
for the study of complex phenomena.  Some systems can self organize
with physics at many scales, while others provide fascinating
demonstrations of thermodynamic laws.  I have only touched on a few
issues here, leaving out many related topics such as lattice gasses,
driven interfaces in random media, growth processes, and evolution.
As the ease of programming and the speed of modern computers continue
to rush forward, so will the fascination with such models.

\section*{Acknowledgments} 
I am thankful for discussions with many people, but most particularly
P. Bak, D. Dar, E. Fredkin, N. Margolis, M. Paczuski, T. Toffoli,
F. Van Scoy, and G. Vichniac.  This manuscript has been authored under
contract number DE-AC02-76CH00016 with the U.S.~Department of Energy.
Accordingly, the U.S.~Government retains a non-exclusive, royalty-free
license to publish or reproduce the published form of this
contribution, or allow others to do so, for U.S.~Government purposes.

\end{document}